\documentclass[twoside,slac_one]{revtex4}
\usepackage{graphicx}
\usepackage{fancyhdr}
\usepackage{amsmath} 
\usepackage{bm}
\usepackage{amsxtra}
\usepackage{amssymb}
\usepackage{amsthm}
\usepackage{latexsym}
\usepackage{lscape}

\begin{document}

\title{MicroBooNE}

%

\author{C. M. Ignarra for the MicroBooNE collaboration}
\affiliation{Massachusetts Institute of Technology, Cambridge, MA, USA}

\begin{abstract}
The MicroBooNE experiment is a 170 ton Liquid Argon Time Projection Chamber (LArTPC) that will begin running at Fermilab in 2013. Its primary physics goal is to explore the low energy excess of events seen by the MiniBooNE experiment and it is the next step in the R\&D to make LAr a viable option for future large neutrino detectors. This talk presented an overview of the MicroBooNE experiment with an emphasis on the light collection system and recent technical advances.
\end{abstract}

\maketitle

\thispagestyle{fancy}

\section{Introduction}
The primary purpose of the MicroBooNE \cite{MicroBooNE} experiment is to explore an anomaly in its predecessor, MiniBooNE, (Sec \ref{sec:miniboone}) by utilizing the unique properties of a LArTPC.  It will also be an important step in the U.S. LArTPC program, being the largest LArTPC to be built in the U.S. to date as well as achieving some important milestones in the technology worldwide.  

MicroBooNE will be located at Fermi National Laboratory along the Booster Neutrino Beam.  The detector hall will be located slightly upstream of the MiniBooNE detector hall and below ground on the beamline.  MicroBooNE will contain 170 tons of LAr with a fiducial volume of roughly 60 tons.

\section{LArTPCs}
LArTPCs consist of a volume of LAr with an applied electric field produced by a potential difference across the detector (125 kV, or 500 V/cm in MicroBooNE), and planes of wires on the $V=0$ end of this field.  There are also often phomultiplier tubes (PMTs) to see the scintillation light from the argon.  

\begin{figure*}[t]
\centering
\includegraphics[width=135mm]{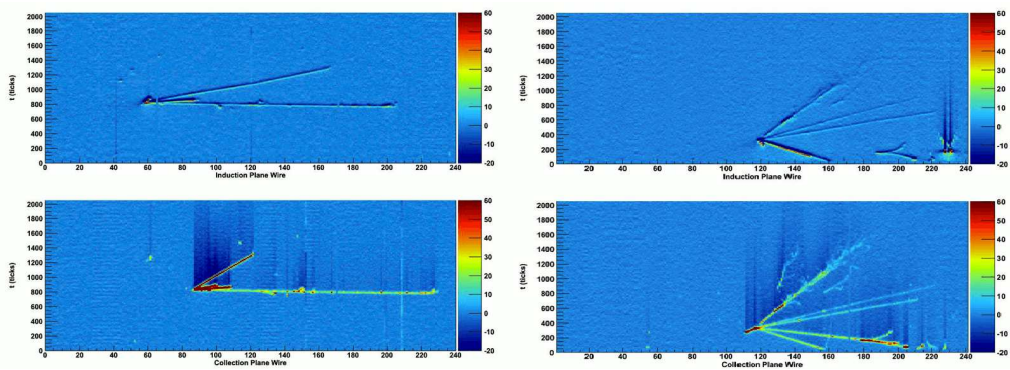}
\caption{Two example event displays from the ArgoNeuT experiment.  ArgoNeuT was a 170 L LArTPC located along the NUMI beamline that was used to measure neutrino cross-sections on argon.  It was also an R\&D project for LArTPC technology.} \label{fig:eventdisplay}
\end{figure*}

When a charged particle travels through the argon, it ionizes the argon atoms, producing ionization electrons along its path.  The ionization electrons rapidly reach terminal velocity and drift towards the wire planes due to the applied electric field.  MicroBooNE has 3 wire planes--the first two are induction planes, which measure the induced current in the wires by the ionization electrons, and the third is a collection plane, which collects these electrons.  The information from these planes allows the 3D paths of the initial charged particles traveling through the detector as well as their deposited energy to be reconstructed (Figure \ref{fig:eventdisplay}).  From this, we can reconstruct the energy of the neutrino that produced these charged particles.

\section{Motivation}


\subsection{Primary motivation}\label{sec:miniboone}

\begin{figure}[]
\centering
\includegraphics[width=100mm]{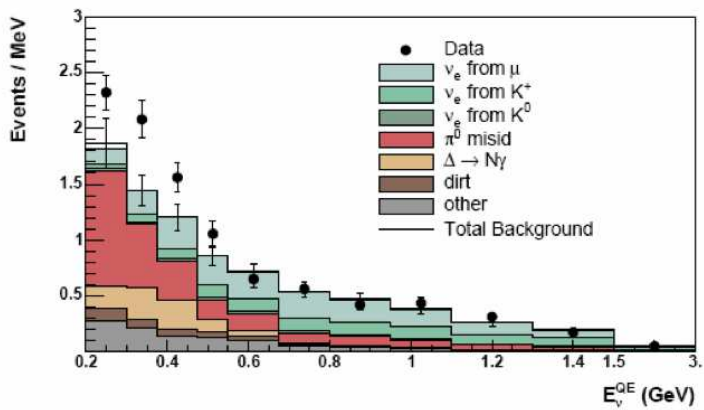}
\caption{MiniBooNE $\nu_e$ candidate data as a function of energy.  An excess of events above background is visible at low energies.} \label{fig:miniboone}
\end{figure}

The primary motivation for MicroBooNE is an anomalous result seen in the MiniBooNE experiment.    MiniBooNE is a Cherenkov detector searching for $\nu_\mu\rightarrow\nu_e (\bar{\nu}_\mu\rightarrow\bar{\nu}_e)$ neutrino oscillations.  MiniBooNE measured an unexplained 3$\sigma$ excess of events at low energies (Fig \ref{fig:miniboone}).   This excess can be due to some unexpected background or new physics.  The main problem with determining the nature of this excess is that Cherenkov detectors cannot distinguish converting photons ($\gamma \rightarrow e^+ e^-$) from electrons, since Cerenkov detectors only have a ~20cm vertex resolution where $<$ 2cm or the ability to measure $dE/dx$ would be needed to distinguish $\gamma$ s from $e^-$s.

\begin{figure}[h]
\centering
\includegraphics[width=80mm]{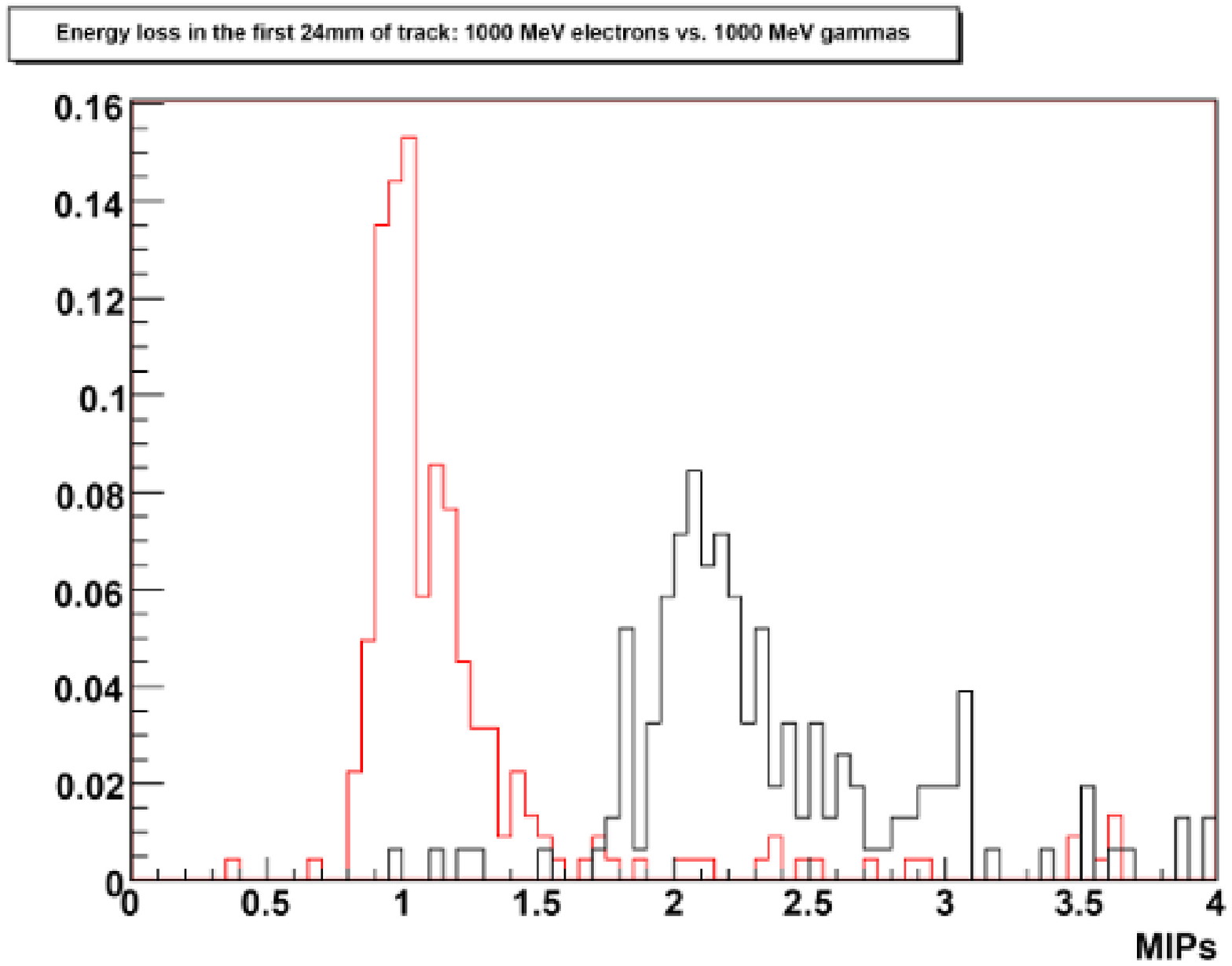}
\caption{Energy loss in the first 24mm of track.  Left: 1 GeV electrons, Right: 1 GeV photons ($\gamma \rightarrow e^+ e^-$)} \label{fig:egamma}
\end{figure}

MicroBooNE however, is sensitive to the different amounts of energy they deposit and can produce a high resolution reconstruction of events.  There is a clear separation in MicroBooNE between electrons, which deposit ~1 MIP of energy, and photons converted into electron-positron pairs, which deposit ~2 MIPs of energy in the first few centimeters of their tracks (Fig \ref{fig:egamma}).  

\subsection{Liquid argon R\&D}
MicroBooNE is also an important stage in the development of LAr technology in the community.  MicroBooNE will be the largest LArTPC to be built in the U.S.  It will also be the first LArTPC to attempt to drift ionization electrons 2.5m.  Demonstrating this 2.5m drift is an important milestone which must be achieved in order to be able to build larger LArTPCs.  

The goal of our LArTPC R\&D is to make this technology a viable option for next generation kiloton scale detectors.  LArTPC technology has many features that can benefit neutrino physics, but we first must demonstrate the scalability of the technology.  Some of the challenges that MicroBooNE will be addressing are noise-reducing cold electronics, LAr purity in larger detectors, analysis tools, and understanding cross sections in LAr.  ArgoNeuT \cite{argoneut} has already made important measurements of cross sections in LAr, but MicroBooNE will build on this with its larger fiducial volume and higher statistics.


\subsection{Other physics goals}

MicroBooNE has a variety of other physics goals as well.  Firstly, the low energy neutrino cross section measurements are an important physics goal.  It is important to understand how to model neutrino interactions on different nuclei.  MicroBooNE is also sensitive to a burst supernova.  It is sensitive to all neutrino species for elastic scattering, charged current, and neutral current events.  This is not unique to LAr, but we can learn a lot from detecting these neutrinos in LAr, particularly in future larger detectors.

We are also able to prepare for future proton decay searches ($p^+ \rightarrow K^+\nu$).  MicroBooNE is not large enough to see proton decay, but we can develop particle ID, triggers, and understand background.  The ability to see proton decay is a very attractive feature of LAr detectors as the outgoing K is below Cherenkov threshold and therefore invisible to Cherenkov detectors, which are currently the leading detector type for kiloton scale detectors.

MicroBooNE is also sensitive to $\Delta S$, the fraction of proton spin carried by the strange quark, through neutral current elastic scattering.  This is impossible for most detectors because it is difficult to tell protons from neutrons, but liquid argon detectors can measure the energy of the outgoing proton and may be able to see the disconnected neutron-proton vertex.  Measuring $\Delta S$ in LAr will help us to better understand proton spin.

Lastly, MicroBooNE will be able to search for exotic physics such as decays of exotic heavy particles.


\section{Light in LAr}


\subsection{Importance of light detection}

While the information provided by the argon scintillation light is not directly used in the determination of the energy of the incident neutrinos, the $t_0$ it provides is required.  Without it, there is some uncertainty between where the event occurred in the detector and when the event took place, since we only know when the event reached the wire planes.   A light collection system allows us to reject background by comparing interaction time with the beam time structure, which is crucial for a surface detector like MicroBooNE.  It also allows us to trigger on interesting non-beam events, which is necessary in order to study events like from supernovae or proton decay in liquid argon.  Knowing the exact location of an event can also allow corrections for charge losses and diffusions as a function of drift distance which allows for a more accurate measurement of energy deposits.


\subsection{Light Production in LAr} \label{sec:light_in_lar}
The light that we are interested in measuring is the scintillation light of the argon.  There are two paths for scintillation light to be produced \cite{LArScint} \cite{LarScint2}:

\begin{description}
\item[The fast scintillation path] makes up about 25\% of the scintillation light.  As a charged particle travels through the LAr, it may excite the argon atoms.  When this occurs, the excited argon atom may combine with another argon atom to form a singlet state excimer which will then decay to form two argon atoms and a 128nm photon.  This process takes ~ 6ns.
\item[The slow scintillation path] makes up the other 75\% of the scintillation light and begins with the charged particle traveling through the LAr ionizing the argon atom (the same process by which the ionization electrons are produced).  The resulting argon ion can recombine with another argon atom and another electron in order to produce a triplet state excimer.  This excimer then decays into a singlet state excimer, which as before, decays into two argon atoms and a 128 nm photon.  This process takes 1.6 $\mu s$
\end{description}

This 128 nm scintillation light is in the vacuum UV, a range of wavelengths which is unable to pass through most substances, including air.   This 128 nm light is also unable to pass through the glass of a photomultiplier tube, so we must first shift the light to wavelengths in the visible range in order to detect it.


\subsection{Wavelength shifting}

\begin{figure}[h]
\centering
\includegraphics[width=80mm]{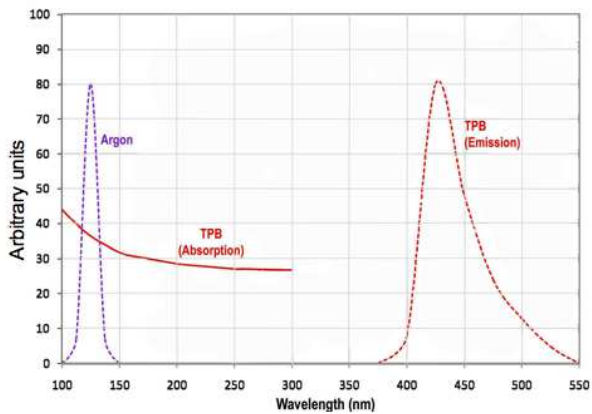}
\caption{A plot of the TPB absorption and emission spectra, overlaid with the argon scintillation spectrum.  The TPB emission spectrum is a good match to the quantum efficiency of our PMTs.} \label{fig:tpbspectrum}
\end{figure}

In order to detect the argon scintillation light with photomultiplier tubes (PMTs), we use a wavelength shifting material called Tetraphenyl Butadiene (TPB) to coat acrylic plates positioned in front of the PMTs.  TPB absorbs light in the UV and re-emits in the visible (Fig \ref{fig:tpbspectrum}).  The efficiency for re-emission of light absorbed at 128nm has been determined by ref \cite{vick} to be 120\% for a film of pure TPB made through vacuum evaporation.  We plan to use a coating of 50\% TPB and 50 \% polystyrene (PS) for our plate coating, which we have measured to have about 50\% of the efficiency of  the evaporative coating. We find that this mixture makes the plates more durable and also more cost effective.  

\subsection{Light collection system}

\begin{figure*}[t]
\centering
\includegraphics[width=135mm]{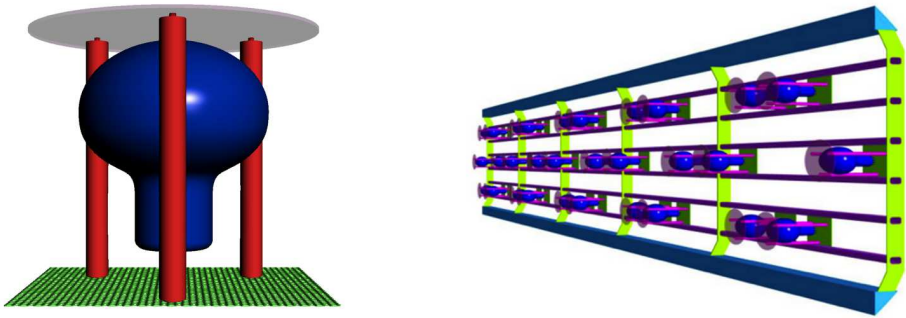}
\caption{A PMT mount displaying the placement of the TPB coated plates above the PMT and the arrangement of PMTs on the support rack in MicroBooNE.} \label{fig:pmtmount}
\end{figure*}

The TPB coated plates have a diameter of 12" and will be placed on top of posts so that they are suspended directly above the PMTs (Fig \ref{fig:pmtmount}).  The posts, which are made of the thermally resistant material polyether ether ketone (PEEK), also serve to hold the PMT in place.  There will be 30 PMTs in MicroBooNE arranged behind the wire planes along the side of the detector outside of the electric field region (Fig \ref{fig:pmtmount}).  

\section{Light Collection R\&D at MIT}
While the MicroBooNE light collection system is a good fit for a ~100 ton LAr detector, there are many challenges this system would face when scaling up to a larger multi-kiloton detector.  The proposed multi-kiloton detectors use modules which each have their own electric fields and wire planes, so that the electric field does not have to be applied across too large of a distance, as the necessary field strength would be difficult to work with.  Having a light collection system only on the edges of such a detector is unlikely to be successful, as the light has a large probability of being absorbed during multiple Rayleigh scatters.  Having a system that could slide in between wireplanes would be a solution to this problem.

\begin{figure}[h]
\centering
\includegraphics[width=60mm]{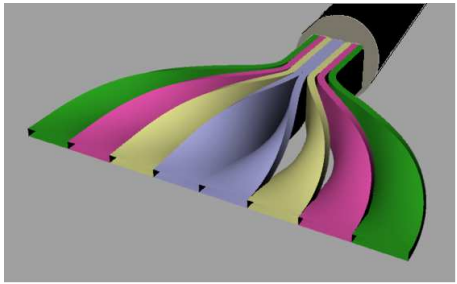}
\caption{An illustration of lightguides forming a paddle and bent so that light is adiabatically transmitted into a PMT.} \label{fig:tesspaddle}
\end{figure}

Our solution is a lightguide detection system \cite{lightguides} where lightguides are bent to guide light adiabatically into PMTs which are placed out of the electric field region.  (see fig \ref{fig:tesspaddle}).  These lightguides would take up far less of the fiducial volume of the detector than a system of PMTs, as well as being able to go places that PMTs cannot such as in electric fields and tighter spaces.  A much larger area of coverage is also possible.

\subsection{Acrylic Rods as Light Guides}
In order for lightguiding to occur without the light having originally entered at the end of the guide, the light must be produced inside the guide.  This is accomplished by embedding the TPB in a thin film of polystyrene (PS), a good index of refraction match to acrylic, and applying this coating to the surface of the acrylic.  The coating is a mixture of 25\% TPB to 75\% PS.  When the UV light hits a TPB molecule, the outgoing visible light is emitted isotropically inside the film and may be totally internally reflected if the light is emitted above the critical angle of the guide. This coating must be optically smooth for better lightguiding properties.


\subsection{Testing}
\begin{figure}[h]
\centering
\includegraphics[width=50mm]{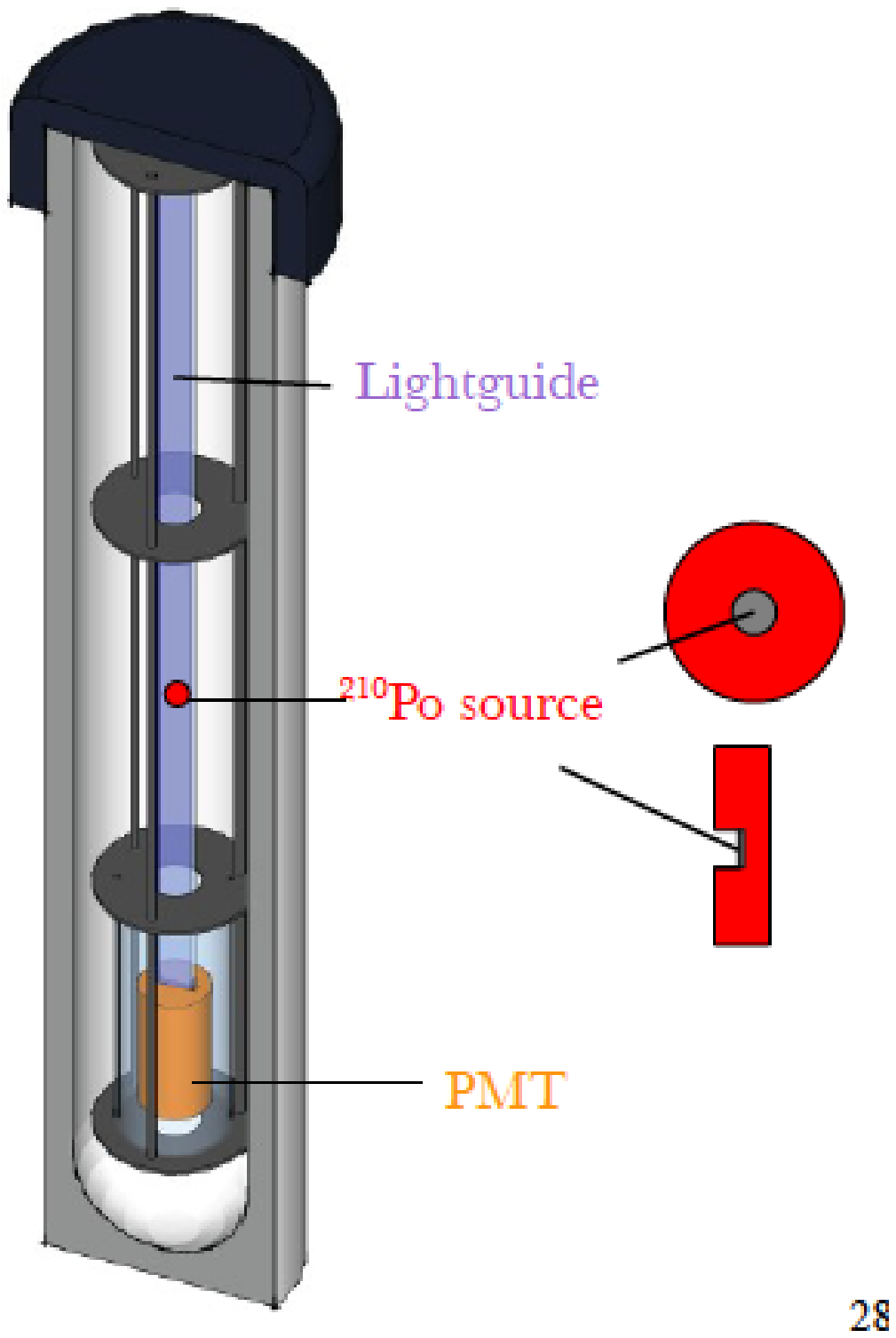}
\caption{Lightguide test setup.  A  $^{210}Po$ disk source placed $~ $5mm from the lightguide emits photons through argon scintillation which hit the lightguide, are shifted, and reemitted down the lightguide and into the PMT} \label{fig:lightguide_setup}
\end{figure}

\begin{figure}[h]
\centering
\includegraphics[width=80mm]{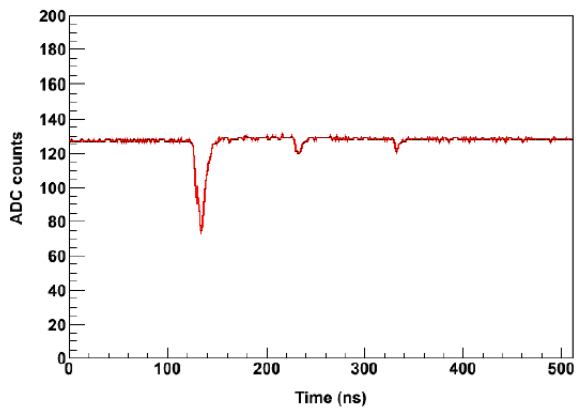}
\caption{An example pulse from a phototube using our setup.  Both prompt and late pulses are visible. An AlazarTech ATS9870 digitizer is used for data aquisition.} \label{fig:pulse}
\end{figure}

The lightguides are tested using a small $^{210}Po$ disk source placed $\sim$ 5 mm from the lightguide which produces $5.3$ MeV $\alpha$ particles.  These $\alpha s$ travel $50 \mu m$ in argon, producing photons along its path through the mechanism described in section \ref{sec:light_in_lar} which may hit the lightguide and be transmitted to a R7725-mod PMT.  Our setup is shown in Fig \ref{fig:lightguide_setup}. The PMT observes a multi-photoelectron prompt pulse from the fast scintillation light, which arrives all at once, as well as single photoelectron (p.e.) late pulses (Fig \ref{fig:pulse}).

These pulses are calibrated in two ways in ref \cite{lightguides}.  We must define one p.e. in terms of ADC counts.  The first method is to define one p.e. by the late light, as most of the late light pulses only contain one p.e., and take the mean of the gaussian that results by fitting the pulse heights of these late light charges.  The second method is to fit the early pulse with equally spaced Gaussians (with the spacing allowed to float).  There are natural peaks in the data, representative of a Poisson distribution around each number of p.e. that  a pulse can have.  Both of these methods yield the same answer.  For our new studies, we are using the integrated pulse charge and calibrate using the late light only for simplicity.

The resulting fits \cite{lightguides} yield an average of ~8 p.e. and an attenuation of ~75 cm.  Using 1 m paddles (8 lightguides curving into a 2 in PMT), we would need 27 paddles to detect a 40 MeV proton (defined as observing 5 p.e.) in MicroBooNE.  This is practical, though not nearly as good as the current MicroBooNE system.


\subsection{New Developments}
Since this paper, we have switched to using UV transmitting acrylic for our coating in which to embed the TPB, and switched to using cast acrylic instead of extruded acrylic for our lightguides, as the cast acrylic should have a higher attenuation length.  With these new lightguides, we are seeing an average of 35 p.e. per $\alpha$, which is a factor of 4 more light.   This makes this R\&D even more promising.  

\begin{figure}[h]
\centering
\includegraphics[width=80mm]{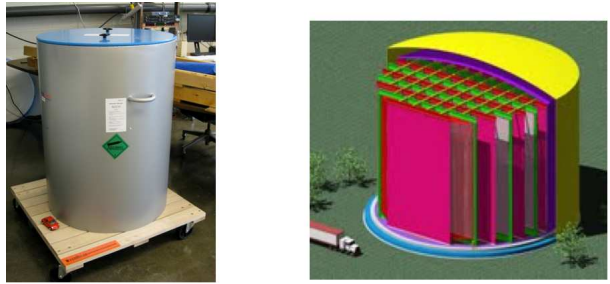}
\caption{Left: The dewar which will become our demonstration detector. There will be 6 paddles consisting of 8 1m long lightguides each running vertically along the sides of the detector.  Right:  Next Generation multi-kiloton LAr detector.} \label{fig:demodetector}
\end{figure}

We plan to set up a demonstration detector (Fig \ref{fig:demodetector}. with six 20 cm wide by 40 cm long paddles of eight lightguides each in order to demonstrate this technology as well as measure the muon lifetime in argon.

\section{Conclusion}

MicroBooNE is an important state in the development of future LAr TPCs.  It will also help our understanding of the MiniBooNE experiment as well as making other important physics contributions.  The new developments in lightguide R\&D for general purpose light collection in LAr look promising, and may be a solution to the problems that will be faced when scaling up systems which are currently in use to the kiloton scale.


\end{document}